\documentclass[aps,pre,twocolumn,groupedaddress,10pt]{revtex4-1}

\usepackage[active]{srcltx}
\usepackage{graphicx}
\usepackage{amsmath, amssymb, amsfonts}
\usepackage{color}

\begin{document}


\title{Random walk in degree space and the time-dependent Watts-Strogatz model}


\author{H. L. Casa Grande$^{1,a}$, M. Cotacallapa$^{1, 2}$, M. O. Hase$^{1}$}
\affiliation{$^{1}$ Escola de Artes, Ci\^encias e Humanidades, Universidade de S\~ao Paulo, Av. Arlindo B\'ettio 1000, 03828-000 S\~ao Paulo, Brazil\\
$^{2}$ Instituto Nacional de Pesquisas Espaciais, 12227-010, S\~ao Jos\'e dos Campos, S\~ao Paulo, Brazil}



\begin{abstract}
In this work, we propose a scheme that provides an analytical estimate for the time-dependent degree distribution of some networks. This scheme maps the problem into a random walk in degree space, and then we choose the paths that are responsible for the dominant contributions. The method is illustrated on the dynamical versions of the Erd\H{o}s-R\'enyi and Watts-Strogatz graphs, which were introduced as static models in the original formulation. We have succeeded in obtaining an analytical form for the dynamics Watts-Strogatz model, which is asymptotically exact for some regimes.
\end{abstract}

\pacs{89.75.Hc, 02.50.Ga, 02.50.Ey}
\email{helder@if.usp.br}



\maketitle



\section{Introduction}

The investigation of structure and dynamics of networks has been a powerful strategy to analyze interacting many-body problems present in many different areas: biological, ecological, economical and social systems, to name some of them. The map of these systems into graphs is a fruitful old idea, and the knowledge of the interconnection between its vertices is a necessary condition that allows us to examine a myriad of pratical problems \cite{AB02, DM03, N03, BLMCH06, NB06, N10}.

Nowadays, there are several research interests involving complex networks. We can say, for instance, that there is an effort to obtain a better understanding of networks from some of its internal structures like the formation of communities \cite{GN02, NG04}, or a more complex interconnection of graphs like the multilayer networks \cite{BBCdGGGRSNWZ14, KABGMP14}. The complexity of the internal structure reflects on the entropy of the network \cite{B08, AB09}, which shows the possibility of classifying several internal structures, and, as an application, it is possible to assess information of the vertices of a network by an inference approach through measuring its entropy \cite{BPM09}. At the same time, we still have progress on important questions that use complex networks as a framework to define other problems on it; for instance, we can cite the active area of epidemiological models \cite{PSCVMV15}, or statistical models on complex networks to analyze critical phenomena \cite{DGM08}.

At this point, it is worth mentioning that despite the progress in several directions, it is natural that analytical results are less frequent than numerical ones, which is understandable due to the technical complexities presented by many relevant questions. Furthermore, many existing analytical results come from stationary regime. In this scenario, we propose a scheme that estimates the time-dependent degree distribution. In order to illustrate our idea, we revisited the Watts-Strogatz model \cite{WS98}. Although not being a \textquotedblleft complex network\textquotedblright in the sense that it does not display a heterogeneous degree distribution \cite{BA99}, it has small-world property and has high clustering \cite{WS98, BW00}, two properties shared with many real networks. The model was originally defined as an intermediate configuration between a regular lattice and a graph where all their nodes are randomly linked, and we will present a slightly modified version from the original one in order to capture its dynamical evolution analytically.

The layout of this work is as follows. In purpose of illustrating the main idea of the work, we start with a dynamical version of the Erd\H{o}s-R\'enyi model \cite{ER59, G59} in section 2 and we introduce the main model, the time-dependent Watts-Strogatz graph, in section 3. Then, we present the main idea that allow one to achieve an analytical form for the dynamic degree distribution in section 4. Some final comments are presented in section 5.


\section{Time-dependent Erd\H{o}s-R\'enyi model}
\label{tder}

The initial condition of the model consists of $N$ vertices and no edges at time $t=0$. At each time step, two vertices are randomly chosen and linked; this includes the possibility of having a loop (an edge that connects a vertex to itself). It is clear that each end of an edge links to a vertex with probability $1/N$. Therefore, defining $p(k,s,t)$ as the probability that a vertex $s$ has degree $k$ at time $t$, one can represent the dynamics as
\begin{align}
\nonumber \lefteqn{p(k,s,t+1) = w_{\textnormal{ER}} (k|k-2)p(k-2,s,t) +}& \\
 &+ w_{\textnormal{ER}} (k|k-1)p(k-1,s,t) + w_{\textnormal{ER}} (k|k)p(k,s,t)\,,
\label{er_me}
\end{align}
with $p(k,s,t=0)=\delta_{k,0}$ as the initial condition, where $\delta_{k,m}$ is the Kronecker symbol ($\delta_{k,m}=1$ when $k=m$, and $\delta_{k,m}=0$ otherwise). Furthermore, $w_{\textnormal{ER}}(k|m)$ is the time-independent conditional probability of changing the degree of a vertex from $m$ to $k$; in the present case,
\begin{align}
\begin{array}{cl}
w_{\textnormal{ER}} (k|k-2) &= \displaystyle\frac{1}{N^{2}}\,,\\
 & \\
w_{\textnormal{ER}} (k|k-1) &= \displaystyle\frac{2}{N}\left(1-\frac{1}{N}\right)\quad\textnormal{ and }\\
 & \\
w_{\textnormal{ER}} (k|k) &= \displaystyle\left(1-\frac{1}{N}\right)^{2}\,.
\end{array}
\label{W_er}
\end{align}
By introducing the time-dependent degree distribution,
\begin{align}
P(k,t) = \frac{1}{N}\sum_{s=1}^{N}p(k,s,t)\,,
\label{Pkt}
\end{align}
the time evolution equation (\ref{er_me}) can be written as
\begin{align}
\nonumber \lefteqn{P(k,t+1) = \frac{1}{N^{2}}P(k-2,t) +}& \\
 &+ \frac{2}{N}\left(1-\frac{1}{N}\right)P(k-1,t) + \left(1-\frac{1}{N}\right)^{2}P(k,t)\,.
\label{er_meP}
\end{align}
If now one introduces the generating function
\begin{align}
\Phi(K, t) = \sum_{k\geq 0}K^{k}P(k,t)\,,
\label{Z}
\end{align}
the equation (\ref{er_meP}) can be casted as
\begin{align}
\nonumber\lefteqn{\Phi(K,t+1) =}& \\
 &= \Bigg[\frac{K^{2}}{N^{2}} + \frac{2K}{N}\left(1-\frac{1}{N}\right)+ \left(1-\frac{1}{N}\right)^{2}\Bigg]\Phi(K,t)\,.
\label{er_mePhi}
\end{align}
Introducing, now, the operator
\begin{align}
\mathcal{L}^{\textnormal{ER}}:=\frac{K^{2}}{N^{2}} + \frac{2K}{N}\left(1-\frac{1}{N}\right) + \left(1-\frac{1}{N}\right)^{2}\,,
\label{L_er}
\end{align}
it possible to see that
\begin{align}
\nonumber\Phi(K,t) &= \mathcal{L^{\textnormal{ER}}}\Phi(K,t-1) \\
 &=\left(\mathcal{L^{\textnormal{ER}}}\right)^{2}\Phi(K,t-2) = \cdots = \left(\mathcal{L^{\textnormal{ER}}}\right)^{t}\Phi(K,0)\,,
\label{PhiPhi0}
\end{align}
where the initial condition is $\Phi(K,0)=1$. Therefore, since $\mathcal{L}^{\textnormal{ER}}=\left(\frac{K}{N}+1-\frac{1}{N}\right)^{2}$, one has
\begin{align}
\nonumber\Phi(K,t) &= \left(\frac{K}{N}+1-\frac{1}{N}\right)^{2t} \\
 &= \sum_{m=0}^{2t}{2t\choose m}\left(1-\frac{1}{N}\right)^{2t-m}\left(\frac{K}{N}\right)^{m}\,.
\label{PhiPhiK}
\end{align}
One sees that the right-hand side of (\ref{PhiPhiK}) is a polynomial in $K$, and the time-dependent degree distribution $P(k,t)$ is the coefficient of the term of order $K^{k}$ (which we will refer as \textquotedblleft $K^{k}$-term\textquotedblright) in the right hand side of (\ref{PhiPhiK}). Hence, by a direct inspection, the time-dependent degree distribution is
\begin{align}
P(k,t) = {2t\choose k}\left(1-\frac{1}{N}\right)^{2t-k}\frac{1}{N^{k}}\,,
\label{Pkt_er}
\end{align}
which is a binomial distribution with parameters $2t$ (number of trials) and $1/N$ (success probability in each trial). This result will be revisited in section \ref{rw}, where we will treat the problem of finding the time-dependent degree distribution as a random walk in degree space.

When $t=N\left(N-1\right)/2$, which is the time equivalent to the number of possible distinct edges, one recovers the usual Poisson distribution from the binomial distribution (\ref{Pkt_er}) for $N\gg 1$,
\begin{align}
P(k,t) \simeq \frac{1}{k!}\left(\frac{2t}{N}\right)^{k}e^{-\frac{2t}{N}} \quad\textnormal{ and }\quad\langle k\rangle = \frac{2t}{N}\,.
\label{Poisson}
\end{align}

The exact form of the time-dependent degree distribution can be used to investigate the Shannon entropy,
\begin{align}
\nonumber S(t) &= -\sum_{k}P(k,t)\ln P(k,t) \\
\nonumber &= -\sum_{k=0}^{2t}{2t\choose k}\left(1-\frac{1}{N}\right)^{2t-k}\frac{1}{N^{k}}\ln{2t\choose k} + \\
 &+ 2t\left[\frac{1}{N}\ln\left(N-1\right) - \ln\left(1-\frac{1}{N}\right)\right]\,,
\label{shannon}
\end{align}
where the last term is the part of $\ln P$ that could be averaged over the degree distribution trivially.

The profile of the entropy can be investigated numerically and is presented in figure 1. It starts from a low value and achieves the maximum for $t\approx N^{2}/2$, which is when the original (static) Erd\H{o}s-R\'enyi model realizes, and the inclusion of more connections decreases the entropy, as one can see from (\ref{Poisson}). This phenomenon can be heuristically understood by realizing that the inclusion of edges randomly (with uniform probability to each possible pair of nodes) leads the distribution to converge to a Kronecker delta, \textit{i.e.}, the vertices tend to have all the same degree (that increases with time) from the statistical standpoint.

\begin{center}
\includegraphics[width=245pt]{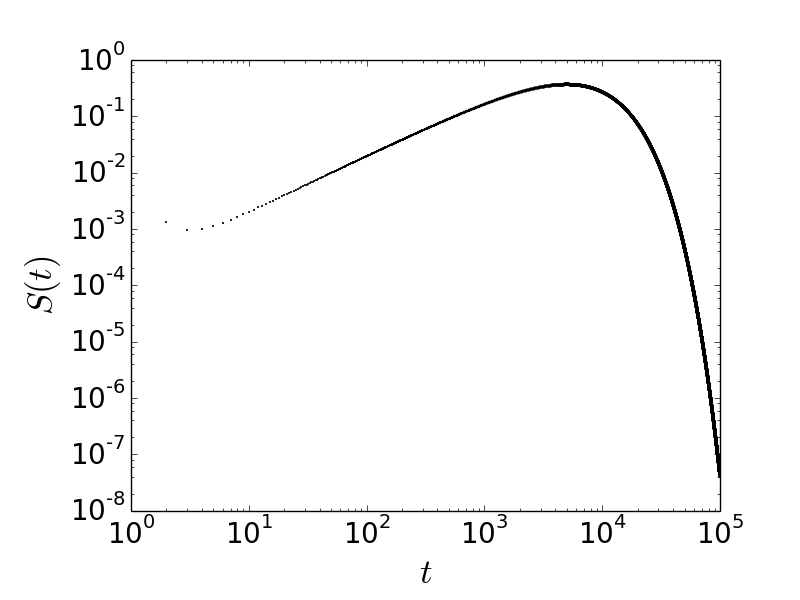}
\end{center}
Figure 1: Entropy of time-dependent Erd\H{o}s-R\'enyi model ($N=100$ and $p=0.01$).


\section{Time-dependent Watts-Strogatz model}
\label{tdws}

The Watts-Strogatz model \cite{WS98} is a small-world network that, unlike the Erd\H{o}s-R\'enyi graph, keeps high clustering. The analytical approach treats it as a static model, despite the fact that it is obtained as an intermediate configuration in rewiring process between a regular lattice and a random graph. We will define a dynamical model that generates a small-world network similar to the one introduced by Watts and Strogatz. Although being slightly different from the original Watts-Strogatz model, it is statistically equivalent and suitable for analytical investigations.

The initial condition of our model consists of a ring with $N$ vertices, and each vertex has degree $k_{0}$ by having a single link to its $k_{0}/2$ next-neighbors as in Watts-Strogatz model. The model has, therefore, $k_{0}N/2$ edges with total degree $M=k_{0}N$. The dynamics obeys the following scheme:

\medskip
\noindent
(i) An edge end is chosen with uniform probability $\frac{1}{M}$.

\medskip
\noindent
(ii) This extremity is reconnected with probability $p$ (and kept without reconnection with probability $1-p$).

\medskip
\noindent
(iii) Back to (i) (repetition for a fixed number of iterations).

\medskip

Therefore, the probability $p(k,s,t)$ of a vertex $s$ having degree $k$ at time $t$ obeys the discrete time recurrent equation
\begin{align}
\nonumber\lefteqn{p(k,s,t+1) = w(k|k-1)p(k-1,s,t) +}& \\
 &+ w(k|k+1)p(k+1,s,t) + w(k|k)p(k,s,t)\,,
\label{me}
\end{align}
where $w(k|m)$ stands for the discrete-time transition rate (conditional probability) from the state of degree $m$ to degree $k$, as in the previous section. Furthermore, the initial condition is $p(k,s,t=0)=\delta_{k,k_{0}}$

Consider now a vertex $s$ at time $t$; it can have degree $k$ at time $t+1$ in the following scenarios:

\medskip
\noindent
I) The vertex $s$ has degree $k-1$ at time $t$ and degree $k$ at time $t+1$: an edge-end, which is not connected to $s$, is chosen with probability $1-\frac{k-1}{M}$. Then, it rewires with probability $p$, and links to $s$ with probability $\frac{1}{N}$; therefore, one has
\begin{align}
w(k|k-1) = \frac{p}{N}\left(1-\frac{k-1}{M}\right)\,.
\label{w(k|k-1)}
\end{align}

\medskip
\noindent
II) The vertex $s$ has degree $k+1$ at time $t$ and degree $k$ at time $t+1$: an edge-end connected to $s$ is chosen with probability $\frac{k+1}{M}$. Then, it rewires with probability $p$, and links to another vertex, say $s^{\prime} (\neq s)$, with probability $1-\frac{1}{N}$; therefore, one has
\begin{align}
w(k|k+1) = \frac{k+1}{M}p\left(1-\frac{1}{N}\right)\,.
\label{w(k|k+1)}
\end{align}

\medskip
\noindent
III) The vertex $s$ has degree $k$ at time $t$ and remains with degree $k$ at time $t+1$: this scenario is divided in four cases, as follows.

\medskip
\noindent
IIIa) An edge-end connected to $s$ is chosen with probability $\frac{k}{M}$, rewires with probability $p$, and links again to $s$ with probability $\frac{1}{N}$;

\medskip
\noindent
IIIb) An edge-end connected to $s$ is chosen with probability $\frac{k}{M}$, but does not rewire (this happens with probability $1-p$);

\medskip
\noindent
IIIc) An edge-end not connected to $s$ is chosen with probability $1-\frac{k}{M}$, rewires with probability $p$, and links to a vertex that is not $s$ with probability $1-\frac{1}{N}$;

\medskip
\noindent
IIId) An edge-end not connected to $s$ is chosen with probability $1-\frac{k}{M}$, but does not rewire (this happens with probability $1-p$);

\medskip
The conditional probability associated to the union of disjoint events IIIa to IIId is
\begin{align}
\nonumber\lefteqn{w(k|k) = \frac{kp}{MN} + \frac{k}{M}\left(1-p\right) +}& \\
\nonumber &+ \displaystyle p\left(1-\frac{k}{M}\right)\left(1-\frac{1}{N}\right) + \left(1-\frac{k}{M}\right)\left(1-p\right) \\
 &= \displaystyle 1 - \frac{p}{N}\left(1+\frac{kN}{M}-\frac{2k}{M}\right)\,.
\label{w(k|k)}
\end{align}

The dynamics defined above can generate a graph similar to the Watts-Strogatz model. For $t=M$, one has a interval of $p$ where the system displays high clustering and low mean shortest path length, as shown in figure 2.

\begin{center}
\includegraphics[width=245pt]{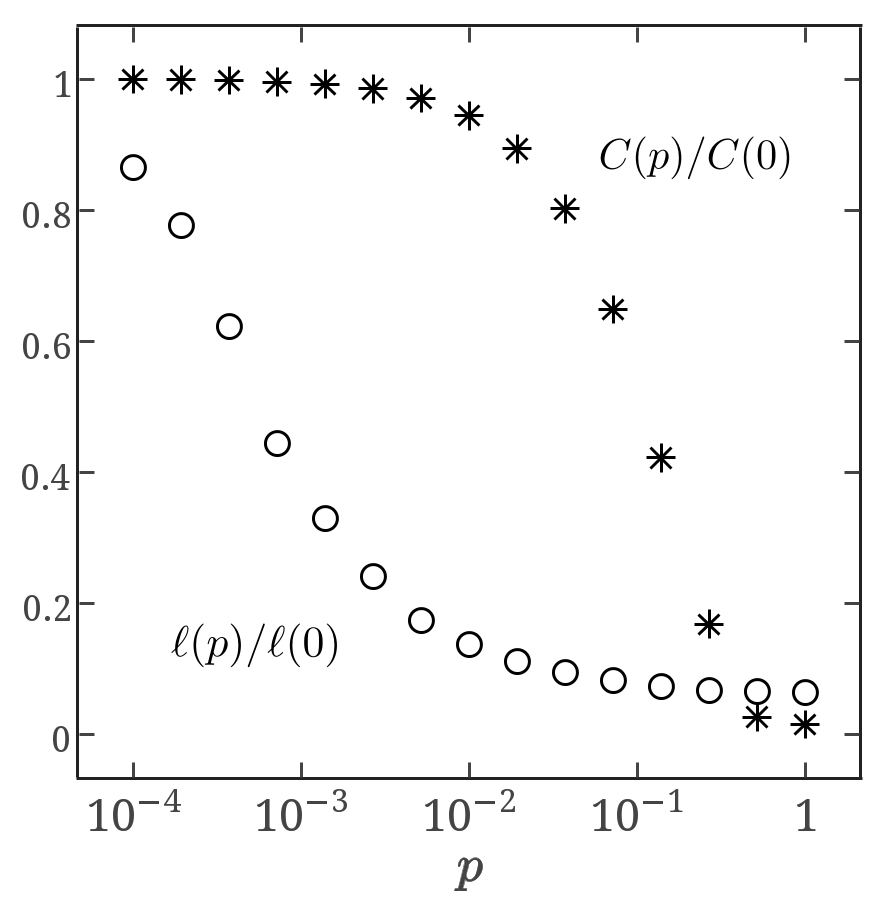}
\end{center}
Figure 2: Clustering $C(p)$ and shortest path length $\ell(p)$ (normalized by $C(0)$ and $\ell(0)$, respectively) of the graph generated by the dynamics of section \ref{tdws}. The parameters are $N=1000$, $k_{0}=10$ and $t=M=k_{0}N$ with $100$ realizations of the simulations; the error bars are smaller than the size of the points.

\bigskip

The time-dependent degree distribution can be evaluated iteratively from the recurrent equation (\ref{me}) and (\ref{Pkt}), and this allows one to compute the entropy $S(t)=-\sum_{k}P(k, t)\ln P(k, t)$ of the model, which is shown in figure 3.

\begin{center}
\includegraphics[width=245pt]{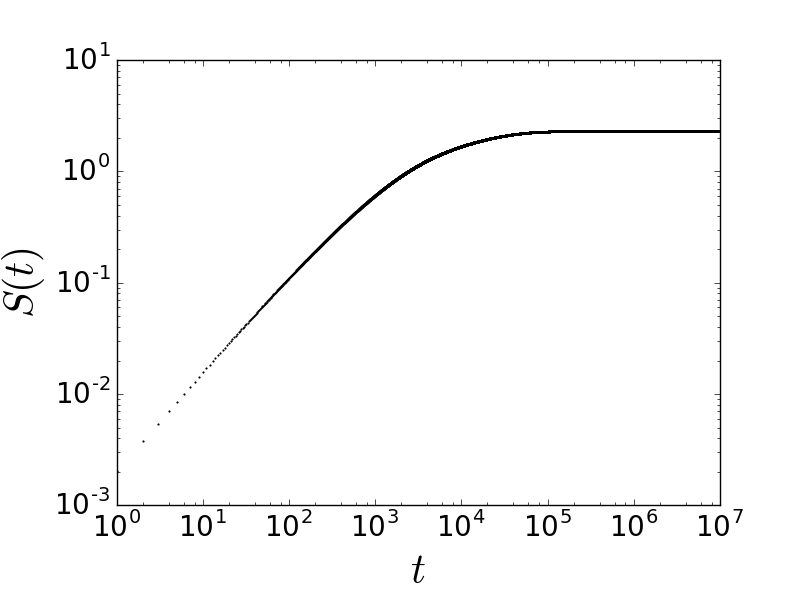}
\end{center}
Figure 3: Entropy of the time-dependent Watts-Strogatz model for $N=100$, $k_{0}=6$ and $p=0.01$.

\bigskip
The entropy starts from a low value, as expected since the initial condition of Watts-Strogatz model is a regular lattice with $P(k,0)=\delta_{k_{0}}$. The entropy, then, grows with time, but reaches a constant value: differently from the Erd\H{o}s-R\'enyi model, the Watts-Strogatz graph has no new connection being added, and the system converges to a stationary degree distribution different from a Kronecker-delta-like as in the Erd\H{o}s-R\'enyi case.

Introducing, again, the generating function (\ref{Z}) to the recurrent equation of the time-dependent degree distribution obtained by combining (\ref{me}) and (\ref{Pkt}), one has
\begin{align}
\Phi(K,t) = \mathcal{L}\Phi(K,t-1) = \mathcal{L}^{t}\Phi(K,0)\,,
\label{Zme}
\end{align}
where the initial condition $\Phi(K,0)=K^{k_{0}}$ stands for each vertex having exactly $k_{0}$ connections. The explicit form of the operator $\mathcal{L}$, which acts on this polynomial, will be presented in the next section. For now, it is sufficient to state that the analytical form of the time-dependent degree distribution is not well explored in the literature.


\section{Random walk in degree space}
\label{rw}

This section is devoted to develop the arguments that will establish analytic results concerning the time-dependent degree distribution of the two models above. The Erd\H{o}s-R\'enyi case will support and illustrate our arguments, since its a simpler laboratory and the exact form (\ref{Pkt_er}) is already known.


\subsection{Time-dependent Erd\H{o}s-R\'enyi model}

As seen in section \ref{tder}, the time-dependent degree distribution $P(k,t)$ is the coefficient of the $K^{k}$-term in $\Phi(K,t)$, as one can see from (\ref{Z}). Moreover, from (\ref{PhiPhi0}) and $\Phi(K,0)=K^{0}=1$, we have $\Phi(K,t)=\left(\mathcal{L}^{\textnormal{ER}}\right)^{t}K^{0}$. This means that one should search for the $K^{k}$-term of a polynomial resulted from the application of $\mathcal{L}^{\textnormal{ER}}$ for $t$ times on $K^{0}$. The operator $\mathcal{L}^{\textnormal{ER}}$, however, can be divided into a sum of three operators, $\mathcal{L}_{0}^{\textnormal{ER}}$, $\mathcal{L}_{1}^{\textnormal{ER}}$ and $\mathcal{L}_{2}^{\textnormal{ER}}$. This separation is convenient, since when these operators are applied on a monomial $K^{m}$ ($m\in\mathbb{Z}$), one has the following behavior:
\begin{align}
\begin{array}{cclccl}
\mathcal{L}_{0}^{\textnormal{ER}}K^{m} &=& \alpha K^{m}\,,& \alpha&:=&\displaystyle\left(1-\frac{1}{N}\right)^{2}\\
 & & & & & \\
\mathcal{L}_{1}^{\textnormal{ER}}K^{m} &=& \beta K^{m+1}\,,& \beta&:=&\displaystyle\frac{2}{N}\left(1-\frac{1}{N}\right)\\
 & & & & & \\
\mathcal{L}_{2}^{\textnormal{ER}}K^{m} &=& \gamma K^{m+2}\,,& \gamma&:=&\displaystyle\frac{1}{N^{2}}
\end{array}\,.
\label{ER_Ldec}
\end{align}
Hence, starting from degree $0$, one can see the procedure of applying $t$ times the operator $\mathcal{L}^{\textnormal{ER}}=\mathcal{L}_{0}^{\textnormal{ER}}+\mathcal{L}_{1}^{\textnormal{ER}}+\mathcal{L}_{2}^{\textnormal{ER}}$ as follows. Since
\begin{align}
\nonumber\lefteqn{\Phi(K,t) =}&\\
 &= \overbrace{\left(\mathcal{L}_{0}^{\textnormal{ER}}+\mathcal{L}_{1}^{\textnormal{ER}}+\mathcal{L}_{2}^{\textnormal{ER}}\right)\cdots\left(\mathcal{L}_{0}^{\textnormal{ER}}+\mathcal{L}_{1}^{\textnormal{ER}}+\mathcal{L}_{2}^{\textnormal{ER}}\right)}^{t\textnormal{ factors}} K^{0}\,,
\label{expl}
\end{align}
the $K^{k}$-term is a sum of many terms, each of them a product of $\mathcal{L}_{0}^{\textnormal{ER}}$, $\mathcal{L}_{1}^{\textnormal{ER}}$ and $\mathcal{L}_{2}^{\textnormal{ER}}$. Let us consider $k=t=2$ as an example; in this case, the $K^{2}$-term of $\Phi(K,2)$ is
\begin{align}
\mathcal{L}_{2}^{\textnormal{ER}}\mathcal{L}_{0}^{\textnormal{ER}}K^{0}+\mathcal{L}_{1}^{\textnormal{ER}}\mathcal{L}_{1}^{\textnormal{ER}}K^{0}+\mathcal{L}_{0}^{\textnormal{ER}}\mathcal{L}_{2}^{\textnormal{ER}}K^{0}\,,
\label{expl2}
\end{align}
and this is $P(k=2, t=2)K^{2}$. In the first term, the system remains with degree zero at time $t=1$ and increases two unities at $t=2$; similar interpretation can be made for the second and third terms. The time-dependent degree distribution is, therefore, a sum of all trajectories, which are random walks in degree space (see figure 4), that leads $k=0$ at $t=0$ to degree $k$ at time $t$. At each time step, the degree can increase one unity, or two unities, or stay constant with probabilities $\beta$, $\gamma$ and $\alpha$, respectively (note that $\alpha+\beta+\gamma=1$ and $\alpha,\beta,\gamma>0$). Hence, denoting by $y_{m}$ the degree at time $m$, it is straightforward that
\begin{align}
\nonumber P(k,t) &= \sum_{\{y_{m}\}}\delta_{y_{0},0}\delta_{y_{t},k}\prod_{m=1}^{t}\big(\alpha\delta_{y_{m}-y_{m-1},0}+\\
 &+\beta\delta_{y_{m}-y_{m-1},1}+\gamma\delta_{y_{m}-y_{m-1},2}\big)\,,
\label{ER_rw}
\end{align}
where $y_{m}\geq 0$ for $0\leq m\leq t$ and the first two Kronecker deltas refer to the initial and final conditions; each term inside the parenthesis indicates if the degree at time $m$ remains constant or increases (with one or two unities) when compared to the degree at the previous instant, $y_{m-1}$.

\begin{center}
\includegraphics[width=245pt]{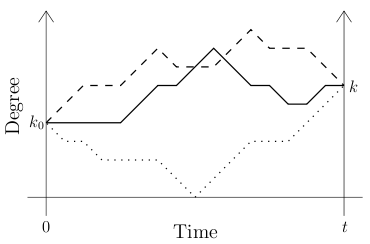}
\end{center}
Figure 4: Three examples of possible evolution of the degree (these examples do not apply for the Erd\H{o}s-R\'enyi model, where the degree never decreses). The initial and final degrees should be $k_{0}$ and $k$, respectively.

\bigskip

The continuous version of (\ref{ER_rw}) is a path-integral formulation of the problem. Nevertheless, it does not lead to an expression that can be trivially tackled by the usual methods.

The time-dependent degree distribution can be evaluated explicitely by exploring the property that $\mathcal{L}_{0}^{\textnormal{ER}}$, $\mathcal{L}_{1}^{\textnormal{ER}}$ and $\mathcal{L}_{2}^{\textnormal{ER}}$ are $c$-numbers. During the time interval $t$, there should be $n_{1}$, $n_{2}$ and $n_{3}$ terms of $\alpha$, $\beta$ and $\gamma$, respectively, such that $n_{1}+n_{2}+n_{3}=t$ and $n_{2}+2n_{3}=k$. Therefore,
\begin{align}
\nonumber\lefteqn{ P(k,t) =}& \\
\nonumber &= \sum_{n_{1},n_{2},n_{3}}\frac{t!}{n_{1}!n_{2}!n_{3}!}\alpha^{n_{1}}\beta^{n_{2}}\gamma^{n_{3}}\delta_{n_{1}+n_{2}+n_{3},t}\delta_{n_{2}+2n_{3},k}\\
 &= \sum_{n=0}^{\left\lfloor\frac{k}{2}\right\rfloor}{t\choose k-n}{k-n\choose n}\gamma^{n}\beta^{k-2n}\alpha^{t-k+n}\,,
\label{ER_step}
\end{align}
which yields the same result of (\ref{Pkt_er}), as expected. In (\ref{ER_step}), $\left\lfloor x\right\rfloor$ is the largest integer equal or less than $x$, and the last equality can be shown after a lengthy induction argument.

Finally, one can also restate the recurrent equation $\Phi(K, t)=\left(\mathcal{L}^{ER}\right)^{t}\Phi(K,0)$ as $\Phi(K, t)=\left(\ell_{1}^{\textnormal{ER}}+\ell_{0}^{\textnormal{ER}}\right)^{2t}\Phi(K,0)$, where now we have two types of operators,
\begin{align}
\ell_{1}^{ER}K^{m}=\frac{1}{N}K^{m+1}\quad\textnormal{ and }\quad\ell_{0}^{ER}K^{m}=\left(1-\frac{1}{N}\right)K^{m}\,,
\label{alternativeL}
\end{align}
that act for an interval of time equal to $2t$ on the initial condition.


\subsection{Time-dependent Watts-Strogatz model}

Similarly as in the previous case, the time-dependent Watts-Strogatz degree distribution is the $K^{k}$-term of $\Phi(K,t)=\mathcal{L}^{t}\Phi(K,0)$, where now the initial condition is $\Phi(K,0)=K^{k_{0}}$ and
\begin{align}
\mathcal{L} := \mathcal{L}_{1} + \mathcal{L}_{0} + \mathcal{L}_{-1}\,,
\label{L}
\end{align}
with
\begin{align}
\begin{array}{cl}
\mathcal{L}_{1} &:= \displaystyle\frac{p}{N}K - \frac{p}{MN}K^{2}\frac{\partial}{\partial K}\\
 & \\
\mathcal{L}_{0} &:= \displaystyle 1 - \frac{p}{N} - \frac{p}{M}K\frac{\partial}{\partial K} + \frac{2p}{MN}K\frac{\partial}{\partial K}\\
 & \\
\mathcal{L}_{-1} &:= \displaystyle\frac{p}{M}\left(1-\frac{1}{N}\right)\frac{\partial}{\partial K}\,.
\end{array}
\label{Lupzerodown}
\end{align}
The form of these operators, which are not $c$-numbers anymore, can be deduced by (\ref{w(k|k-1)}), (\ref{w(k|k+1)}), (\ref{w(k|k)}) and the generating function of (\ref{me}). When these operators are applied on a polynomial of degree $m$, one has
\begin{align}
\begin{array}{ccl}
\mathcal{L}_{1}K^{m} &=& b_{m}K^{m+1}\,,\\
 & & \\
\mathcal{L}_{0}K^{m} &=& a_{m}K^{m}\quad\textnormal{ and}\\
 & & \\
\mathcal{L}_{-1}K^{m} &=& d_{m}K^{m-1}\,,
\end{array}
\label{LLL}
\end{align}
with
\begin{align}
\begin{array}{ccl}
b_{m} &:=&\displaystyle\frac{p}{MN}\left(M-y_{m}\right)\,,\\
 & & \\
a_{m} &:=&\displaystyle 1-\frac{p}{N}-\frac{p}{MN}\left(N-2\right)y_{m} \quad\textnormal{ and}\\
 & & \\
d_{m} &:=&\displaystyle\frac{p}{MN}\left(N-1\right)y_{m}\,.
\end{array}
\label{bad}
\end{align}
Note that now the coefficients $a_{m}$, $b_{m}$ and $d_{m}$ are not constants and the operators $\mathcal{L}_{1}$, $\mathcal{L}_{0}$ and $\mathcal{L}_{-1}$ do not commute as in Erd\H{o}s-R\'enyi case. Following the same argument that has led to (\ref{ER_rw}), we have
\begin{align}
\nonumber P(k,t) &= \sum_{\{y_{m}\}}\delta_{y_{0},k_{0}}\delta_{y_{t},k}\prod_{m=1}^{t}\big(a_{m-1}\delta_{y_{m}-y_{m-1},0}+\\
 &+b_{m-1}\delta_{y_{m}-y_{m-1},1}+d_{m-1}\delta_{y_{m}-y_{m-1},-1}\big)
\label{WS_rw}
\end{align}
for the Watts-Strogatz model. The degree starts with $y_{0}=k_{0}$ at time $t=0$ and ends with $y_{t}=k$ at time $t$. Between these boundaries, the variable $y_{m}$ performs a random walk. This expression is not analytically treatable, and we will invoke some simplifications, which consist of choosing the dominant contributions (paths) to $P(k,t)$.


\subsection{Monotonic paths}

In this section, we will concentrate on the dominant contributions to the degree distribution $P(k,t)$. This follows by choosing a class of paths that starts at $y_{0}=k_{0}$ and ends at $y_{t}=k$. By noticing that $a_{m}=\mathcal{O}(1)\gg b_{m}, d_{m}$, the dominant contributions come from terms that maximize the number of $a_{m}$-factors. This implies minimizing the number of $b_{m}$-factors or $d_{m}$-factors such that they should appear only to change the degree from $k_{0}$ to $k$. In other terms, we have $k-k_{0}$ terms of $b_{m}$ ($d_{m}$) type if $k\geq k_{0}$ ($k<k_{0}$), and the remaining $t-\left(k-k_{0}\right)$ terms are of $a_{m}$ type. Note that these are monotonic paths in the sense that the degree only increases (if $k\geq k_{0}$) or decreases (if $k<k_{0}$).

Let us consider initially, the case $\Delta:=k-k_{0}\geq 0$. Writing the sum of all monotonic paths as being equal to the time-dependent degree distribution leads to
\begin{align}
P(k,t) \approx b_{k_{0}}b_{k_{0}+1}\cdots b_{k-1}\underset{n_{0}+\cdots n_{\Delta}=t-\Delta}{\sum_{n_{0}=0}^{t-\Delta}\cdots\sum_{n_{\Delta}=0}^{t-\Delta}}a_{k_{0}}^{n_{0}}\cdots a_{k}^{n_{\Delta}}\,.
\label{monotonic>}
\end{align}
The $b_{m}$ terms are functions of the degree $y_{m}$ (see equation (\ref{bad})), and not on the instant they appear. In the monotonic crescent path, therefore, each term, $b_{k_{0}},\ldots,b_{k-1}$ should appear one and only one time in this order. The remaining $t-\Delta$ segments of the path are filled by $a_{m}$-terms, and there should be $n_{0}$ of them that are $a_{k_{0}}$, $n_{1}$ of them that are $a_{k_{0}+1}$, and so on (see figure 5). Firstly it is immediate from (\ref{bad}) that
\begin{align}
b_{k_{0}}\cdots b_{k-1} = \left(\frac{p}{MN}\right)^{\Delta}\frac{\left(M-k_{0}\right)!}{\left(M-k\right)!}\,.
\label{bbb>}
\end{align}

\begin{center}
\includegraphics[width=245pt]{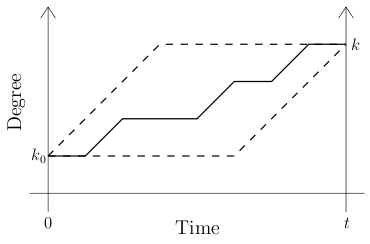}
\end{center}
Figure 5: Increasing monotonic paths. All the monotonic paths, when $k\geq k_{0}$, are located inside the envelope defined by the dashed lines. The upper dashed line corresponds to the path $b_{k_{0}}\cdots b_{k-1}a_{k}^{t-\Delta}$, and the lower dashed line is the monotonic path $a_{k_{0}}^{t-\Delta}b_{k_{0}}\cdots b_{k-1}$.

\bigskip

On the other hand, by using $a_{m}\simeq e^{-\frac{p}{N}-\frac{p}{M}y_{m}}$ one has
\begin{widetext}
\begin{align}
\nonumber\lefteqn{\underset{n_{0}+\cdots n_{\Delta}=t-\Delta}{\sum_{n_{0}=0}^{t-\Delta}\cdots\sum_{n_{\Delta}=0}^{t-\Delta}}a_{k_{0}}^{n_{0}}\cdots a_{k}^{n_{\Delta}} = e^{-\frac{p}{N}\left(t-\Delta\right)}\underset{n_{0}+\cdots n_{\Delta}=t-\Delta}{\sum_{n_{0}=0}^{t-\Delta}\cdots\sum_{n_{\Delta}=0}^{t-\Delta}}e^{-\frac{p}{M}\left(n_{0}k_{0}+\cdots n_{\Delta}k\right)} }&\\
\nonumber &= e^{-\frac{2p}{N}\left(t-\Delta\right)}\sum_{n_{0}=0}^{t-\Delta}e^{-\frac{p}{M}\left(t-\Delta-n_{0}\right)}\sum_{n_{1}=0}^{t-\Delta-n_{0}}e^{-\frac{p}{M}\left(t-\Delta-n_{0}-n_{1}\right)}\cdots\sum_{n_{\Delta-1}=0}^{t-\Delta-n_{0}-\cdots-n_{\Delta-2}}e^{-\frac{p}{M}\left(t-\Delta-n_{0}-\cdots-n_{\Delta-1}\right)} \\
 &= e^{-\frac{2p}{N}\left(t-\Delta\right)}\sum_{u_{0}=0}^{t-\Delta}e^{-\frac{p}{M}u_{0}}\sum_{u_{1}=0}^{u_{0}}e^{-\frac{p}{M}u_{1}}\cdots\sum_{u_{\Delta-1}=0}^{u_{\Delta-2}}e^{-\frac{p}{M}u_{\Delta-1}}\,,
\label{aaa>}
\end{align}
\end{widetext}
where we have performed the change of variables $u_{0}=t-\Delta-n_{0}$, $u_{1}=u_{0}-n_{1}$, $u_{2}=u_{1}-n_{2}$ up to $u_{\Delta-1}=u_{\Delta-2}-n_{\Delta-1}$ in the last passage. Therefore, one has
\begin{align}
\nonumber\lefteqn{\underset{n_{0}+\cdots n_{\Delta}=t-\Delta}{\sum_{n_{0}=0}^{t-\Delta}\cdots\sum_{n_{\Delta}=0}^{t-\Delta}}a_{k_{0}}^{n_{0}}\cdots a_{k}^{n_{\Delta}}\simeq } & \\
 & \simeq e^{-\frac{2p}{N}\left(t-\Delta\right)}\frac{1}{\Delta!}\left(\int_{0}^{t-\Delta}\textup{d}u\,e^{-\frac{p}{M}u}\right)^{\Delta}\,,
\label{aaa>2}
\end{align}
and by (\ref{bbb>}) and (\ref{aaa>2}) one finally finds
\begin{align}
\nonumber P(k,t) &\simeq& \frac{\left(M-k_{0}\right)!}{\left(M-k\right)!}\frac{e^{-\frac{2p}{N}\left(t-\Delta\right)}}{\Delta !}\left[\frac{1-e^{-\frac{p}{M}\left(t-\Delta\right)}}{N}\right]^{\Delta} \\
 & & (k\geq k_{0}, k\in\mathbb{Z})\,.
\label{Pkt>}
\end{align}

The monotonic paths when $k<k_{0}$ is such that
\begin{align}
P(k,t) \approx d_{k_{0}}d_{k_{0}-1}\cdots d_{k+1}\underset{n_{0}+\cdots n_{\Delta}=t-\left|\Delta\right|}{\sum_{n_{0}=0}^{t-\left|\Delta\right|}\cdots\sum_{n_{\Delta}=0}^{t-\left|\Delta\right|}}a_{k_{0}}^{n_{0}}\cdots a_{k}^{n_{\Delta}}\,,
\label{monotonic<}
\end{align}
since now the $d_{m}$-terms are needed to decrease the degree. Since
\begin{align}
d_{k_{0}}\cdots d_{k+1} = \left(\frac{p}{MN}\right)^{\left|\Delta\right|}\left(N-1\right)^{\left|\Delta\right|}\frac{k_{0}!}{k!}\,,
\label{ddd>}
\end{align}
by following a similar procedure as before, one has
\begin{align}
\nonumber P(k,t) &\simeq& \frac{N^{\left|\Delta\right|}k_{0}!}{k!}\frac{e^{-\frac{2p}{N}\left(t-\left|\Delta\right|\right)}}{\left|\Delta\right|!}\left[\frac{1-e^{-\frac{p}{M}\left(t-\left|\Delta\right|\right)}}{N}\right]^{\left|\Delta\right|} \\
 & & (k< k_{0}, k\in\mathbb{Z})\,.
\label{Pkt<}
\end{align}
for $\Delta:=k-k_{0}<0$.

The comparison between the (exact) numerical time-dependent degree distribution obtained from the recurrent equation and the estimations (\ref{Pkt>}) and (\ref{Pkt<}) are shown in figure 6.

The formulas (\ref{Pkt>}) and (\ref{Pkt<}) should be asymptotically exact for $t\ll M$ and $t\simeq M$. The reason for this statement comes from a simple analysis of the order of magnitude of the paths. Remembering that $a_{m}=\mathcal{O}(1)$, $b_{m}=\mathcal{O}(pN^{-1})$ and $d_{m}=\mathcal{O}(pN^{-1})$, a monotonic path is $\mathcal{O}(p^{\left|\Delta\right|}N^{-\left|\Delta\right|})$, while there are $\frac{t!}{\left(t-\left|\Delta\right|\right)!}=\mathcal{O}(t^{\left|\Delta\right|})$ of them. The first correction is due terms that have a $b_{m}$ and $d_{m}$ terms more than the monotonic paths terms (and two $a_{m}$ terms less). Each one of its first correction terms are $\mathcal{O}(p^{\left|\Delta\right|+2}N^{-\left|\Delta\right|+2})$, and there are $\mathcal{O}(t^{\left|\Delta\right|+2})$ of them. The contribution of the first correction is roughly $\mathcal{O}(t^{2}p^{2}N^{-2})$ times the contribution of the monotonic paths. This argument can be extended to corrections of all orders. Therefore, for $p\ll 1$, one expects that the formulas from the monotonic paths only are asymptotically exact for $t\ll M$ and $t\simeq M$. Naturally, the same argument concludes that our estiamtions fail in the case $t\gg M$.

The numerical solution in figure 6 shows that our estimations apply in the case $t\lesssim M$, while the same comment can not be made for $t\gg M$, as expected.

\begin{center}
  \includegraphics[width=245pt]{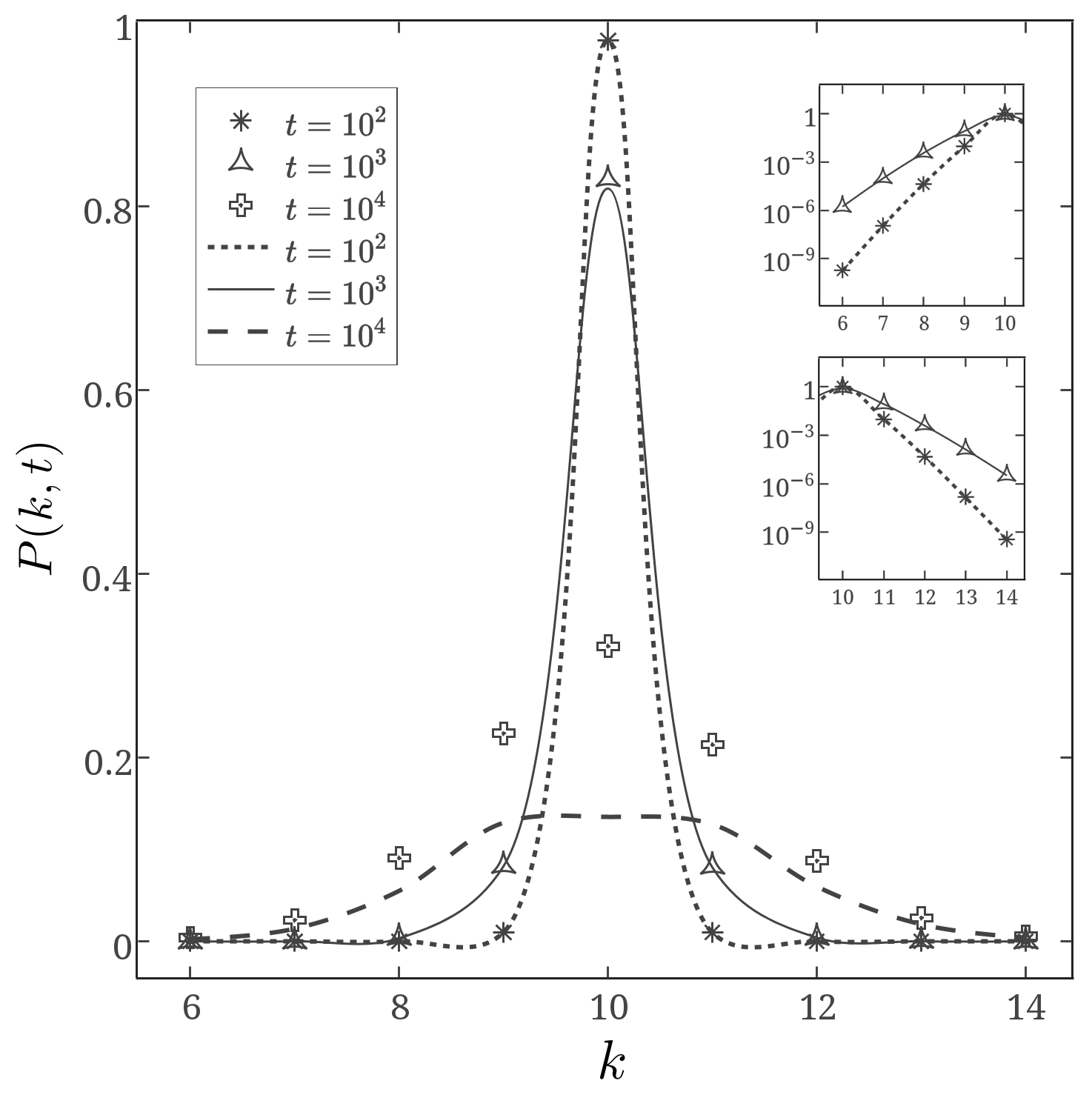}
\end{center}
Figure 6: Time-dependent degree distribution. The points are associated to numerically exact results, and were obtained from the recurrent equation (\ref{me}). The points generated from equations (\ref{Pkt>}) and (\ref{Pkt<}) were interpolated with lines for better visualization. Inset: a detailed visualization of the time-dependent degree distribution (logarithmic scale for the vertical axis) for $t=100$ and $t=1000$.

\bigskip


\section{Conclusion}
\label{conclusion}

In this work, we have formulated the Erd\H{o}s-R\'enyi and Watts-Strogatz graphs as a dynamic model and characterized their behavior from the standpoint of their entropies. We have also examined their time-dependent degree distribution analytically. The Erd\H{o}s-R\'enyi model is analytically accessible, while the same does not extend to the Watts-Strogatz model. We have, nevertheless, obtained a formula that is asymptotically exact for $1\ll t\lesssim M$ and confirmed this validity numerically. The main ideia to achieve this result was to consider the evolution of the degree distribution as a random walk in degree space and select the paths that have dominant contribution. This strategy was specially suitable for networks that has a dynamics which can be written as a recurrent relation like (\ref{PhiPhi0}) or (\ref{Zme}). We have also presented the argument that support the range of validity of our formula, which is based on the estimation of the order of magnitude of contribution of relevant terms.


\section{Acknowledgements}

HLCG thanks PNPD/CAPES (Ed. 82/2014) for financial support. MC acknowledges the OEA scholarship program and CAPES for financial support.


\section{Appendix}

This Appendix is devoted to the proof that equation (\ref{ER_step}) leads to (\ref{Pkt_er}) through the second principle of mathematical induction. Since
\begin{align}
\nonumber P(k,t) &= \sum_{n=0}^{\left\lfloor\frac{k}{2}\right\rfloor}{t\choose k-n}{k-n\choose n}\left[\frac{1}{N^{2}}\right]^{n}\times\\
\nonumber &\times\left[\frac{2}{N}\left(1-\frac{1}{N}\right)\right]^{k-2n}\left[\left(1-\frac{1}{N}\right)^{2}\right]^{t-k+n} \\
 &= \left(1-\frac{1}{N}\right)^{2t-k}\frac{1}{N^{k}}\sum_{n=0}^{\left\lfloor\frac{k}{2}\right\rfloor}{t\choose k-n}{k-n\choose n}2^{k-2n}\,,
\label{ap1}
\end{align}
it remains to show that
\begin{align}
\sum_{n=0}^{\left\lfloor\frac{k}{2}\right\rfloor}{t\choose k-n}{k-n\choose n}2^{k-2n} = {2t\choose k}
\label{ap2}
\end{align}
to complete the proof. As the induction hypothesis, it will be assumed that
\begin{align}
\sum_{n=0}^{\left\lfloor\frac{\kappa}{2}\right\rfloor}{\tau\choose \kappa-n}{\kappa-n\choose n}2^{\kappa-2n} = {2\tau\choose\kappa}
\label{hyp}
\end{align}
is valid for $0\leq\kappa\leq k$ and $0\leq\tau\leq t$. Although the base case is not, in principle, required for the second principle of mathematical induction, we see that both $k=0$ and $k=1$ are satisfied by (\ref{ap2}) for any non-negative $t$ (in particular, $t=0$).

The analysis will be separated in two cases:

\medskip
\noindent
Case (i): Induction on $t$ ($k$ fixed)

\medskip
\noindent
Case (ii): Induction on $k$ ($t$ fixed)

\medskip
Furthermore, the well-known formula
\begin{align}
{\alpha\choose\beta} = {\alpha-1\choose\beta} + {\alpha-1\choose\beta-1}\,,\quad \alpha,\beta\in\mathbb{N}\,,
\label{ap3}
\end{align}
will be extensively invoked, and we take, as usual,
\begin{align}
{\alpha\choose -1}={\alpha\choose\alpha+1}=0\,,\quad\alpha\in\mathbb{N}\cup\{0\}\,.
\label{conv}
\end{align}

\bigskip
\noindent
Case (i): Induction on $t$ ($k$ fixed)

\bigskip
In this case, the left hand side of (\ref{ap2}), for $t\rightarrow t+1$ and $k$ fixed, is
\begin{widetext}
\begin{align}
\nonumber \lefteqn{\sum_{n=0}^{\left\lfloor\frac{k}{2}\right\rfloor}{t+1\choose k-n}{k-n\choose n}2^{k-2n} = \sum_{n=0}^{\left\lfloor\frac{k}{2}\right\rfloor}{t\choose k-n}{k-n\choose n}2^{k-2n} + \sum_{n=0}^{\left\lfloor\frac{k}{2}\right\rfloor}{t\choose k-1-n}{k-n\choose n}2^{k-2n}= }& \\
 &= \sum_{n=0}^{\left\lfloor\frac{k}{2}\right\rfloor}{t\choose k-n}{k-n\choose n}2^{k-2n} + \sum_{n=0}^{\left\lfloor\frac{k}{2}\right\rfloor}{t\choose k-1-n}{k-1-n\choose n}2^{k-2n} + \sum_{n=0}^{\left\lfloor\frac{k}{2}\right\rfloor}{t\choose k-1-n}{k-1-n\choose n-1}2^{k-2n}\,,
\label{ap4}
\end{align}
\end{widetext}
where (\ref{ap3}) was invoked in the first and second passages. By the induction hypothesis (\ref{hyp}), the first term in the last line of (\ref{ap4}) is
\begin{align}
\sum_{n=0}^{\left\lfloor\frac{k}{2}\right\rfloor}{t\choose k-n}{k-n\choose n}2^{k-2n} = {2t\choose k}\,.
\label{ap5}
\end{align}
On the other hand, since $\left\lfloor\frac{k}{2}\right\rfloor=\left\lfloor\frac{k-1}{2}\right\rfloor$ for $k$ odd, and $\left\lfloor\frac{k}{2}\right\rfloor=\left\lfloor\frac{k-1}{2}\right\rfloor+1$ for $k$ even, one has
\begin{align}
\nonumber\lefteqn{\sum_{n=0}^{\left\lfloor\frac{k}{2}\right\rfloor}{t\choose k-1-n}{k-1-n\choose n}2^{k-2n} =}&\\
\nonumber &= \sum_{n=0}^{\left\lfloor\frac{k-1}{2}\right\rfloor}{t\choose k-1-n}{k-1-n\choose n}2^{k-1-2n}\cdot 2 \\
 &= 2{2t\choose k-1}
\label{ap6}
\end{align}
by (\ref{hyp}). The first equality for $k$ even is because the term $n=\left\lfloor\frac{k-1}{2}\right\rfloor+1$ in the sum vanishes due to (\ref{conv}). Finally, by a change of variable, the last term of (\ref{ap4}) can be written as
\begin{align}
\nonumber\lefteqn{\sum_{n=0}^{\left\lfloor\frac{k}{2}\right\rfloor}{t\choose k-1-n}{k-1-n\choose n-1}2^{k-2n} =} & \\
\nonumber &= \sum_{m=0}^{\left\lfloor\frac{k-2}{2}\right\rfloor}{t\choose k-2-m}{k-2-m\choose m}2^{k-2-2m} \\
 &= {2t\choose k-2}
\label{ap7}
\end{align}
by (\ref{hyp}) and using (\ref{conv}). Replacing (\ref{ap5}), (\ref{ap6}) and (\ref{ap7}) in (\ref{ap4}), one has
\begin{align}
\nonumber \lefteqn{\sum_{n=0}^{\left\lfloor\frac{k}{2}\right\rfloor}{t+1\choose k-n}{k-n\choose n}2^{k-2n}= }& \\
 &= {2t\choose k} + 2{2t\choose k-1} + {2t\choose k-2} = {2\left(t+1\right)\choose k}\,,
\label{ap8}
\end{align}
which is the desired result.

\bigskip
\noindent
Case (ii): Induction on $k$ ($t$ fixed)

\bigskip
Let us first restate (\ref{ap2}) as
\begin{align}
\nonumber\lefteqn{\sum_{n=0}^{\left\lfloor\frac{k}{2}\right\rfloor}{t\choose k-n}{k-n\choose n}2^{k-2n} =}& \\
 &= \sum_{n=0}^{\left\lfloor\frac{k}{2}\right\rfloor}{t\choose n}{t-n\choose t-k+n}2^{k-2n} = {2t\choose k}\,,
\label{ap9}
\end{align}
and the last equality will be proved here. Now, the induction hypothesis is
\begin{align}
\sum_{n=0}^{\left\lfloor\frac{\kappa}{2}\right\rfloor}{\tau\choose n}{\tau-n\choose \tau-\kappa+n}2^{\kappa-2n} = {2\tau\choose\kappa}
\label{hyp2}
\end{align}
being valid for $0\leq\kappa\leq k$ and $0\leq\tau\leq t$. In the case (ii), the left hand side of the last line of (\ref{ap9}), for $k\rightarrow k+1$ and $t$ fixed, is
\begin{widetext}
\begin{align}
\nonumber \lefteqn{\sum_{n=0}^{\left\lfloor\frac{k+1}{2}\right\rfloor}{t\choose n}{t-n\choose t-k-1+n}2^{k+1-2n} =} & \\
\nonumber &= \sum_{n=0}^{\left\lfloor\frac{k+1}{2}\right\rfloor}{t-1\choose n}{t-n\choose t-k-1+n}2^{k+1-2n}+\sum_{n=0}^{\left\lfloor\frac{k+1}{2}\right\rfloor}{t-1\choose n-1}{t-n\choose t-k-1+n}2^{k+1-2n} \\
\nonumber &= \sum_{n=0}^{\left\lfloor\frac{k+1}{2}\right\rfloor}{t-1\choose n}{t-1-n\choose t-k-1+n}2^{k+1-2n}+\sum_{n=0}^{\left\lfloor\frac{k+1}{2}\right\rfloor}{t-1\choose n}{t-1-n\choose t-k-2+n}2^{k+1-2n}+ \\
 &+ \sum_{n=0}^{\left\lfloor\frac{k+1}{2}\right\rfloor}{t-1\choose n-1}{t-n\choose t-k-1+n}2^{k+1-2n}\,,
\label{ap10}
\end{align}
\end{widetext}
where (\ref{ap3}) was invoked in the first and second passages. Note that
\begin{align}
\nonumber\lefteqn{\sum_{n=0}^{\left\lfloor\frac{k+1}{2}\right\rfloor}{t-1\choose n}{t-1-n\choose t-k-1+n}2^{k+1-2n} =}& \\
\nonumber &= \sum_{n=0}^{\left\lfloor\frac{k}{2}\right\rfloor}{t-1\choose n}{t-1-n\choose t-1-k+n}2^{k-2n}\cdot 2 \\
 &= 2{2t-2\choose k}\,.
\label{ap11}
\end{align}
In the first passage of (\ref{ap11}), one has $\left\lfloor\frac{k+1}{2}\right\rfloor=\left\lfloor\frac{k}{2}\right\rfloor$ if $k$ is even. If $k$ is odd, the summation ends at $\left\lfloor\frac{k+1}{2}\right\rfloor=\left\lfloor\frac{k}{2}\right\rfloor+1$; however, the term $n=\left\lfloor\frac{k}{2}\right\rfloor+1$ has no contribution to the sum due to (\ref{conv}).

The last term of (\ref{ap10}) can be casted as
\begin{align}
\nonumber\lefteqn{\sum_{n=0}^{\left\lfloor\frac{k+1}{2}\right\rfloor}{t-1\choose n-1}{t-n\choose t-k-1+n}2^{k+1-2n} =}& \\
\nonumber &= \sum_{m=0}^{\left\lfloor\frac{k-1}{2}\right\rfloor}{t-1\choose m}{t-1-m\choose \left(t-1\right)-\left(k-1\right)+m}2^{k-1-2m} \\
 &= {2t-2\choose k-1}
\label{ap12}
\end{align}
by (\ref{hyp2}). Replacing (\ref{ap11}) and (\ref{ap12}) into (\ref{ap10}), and using (\ref{ap3}), one has
\begin{widetext}
\begin{align}
\nonumber\sum_{n=0}^{\left\lfloor\frac{k+1}{2}\right\rfloor}{t\choose n}{t-n\choose t-k-1+n}2^{k+1-2n} &= 2{2t-2\choose k} + {2t-2\choose k-1} +\sum_{n=0}^{\left\lfloor\frac{k+1}{2}\right\rfloor}{t-1\choose n}{\left(t-1\right)-n\choose \left(t-1\right)-k-1+n}2^{k+1-2n} \\
 &= {2t-1\choose k} + {2t-2\choose k} +\sum_{n=0}^{\left\lfloor\frac{k+1}{2}\right\rfloor}{t-1\choose n}{\left(t-1\right)-n\choose \left(t-1\right)-k-1+n}2^{k+1-2n}\,,
\label{ap13}
\end{align}
\end{widetext}
which is a reursive relation in $t$. Therefore, one can write (\ref{ap13}) as
\begin{widetext}
\begin{align}
\nonumber\sum_{n=0}^{\left\lfloor\frac{k+1}{2}\right\rfloor}{t\choose n}{t-n\choose t-k-1+n}2^{k+1-2n} &= {2t-1\choose k} + {2t-2\choose k} + \cdots + {2k+3\choose k} + {2k+2\choose k} + \\
 &+ \sum_{n=0}^{\left\lfloor\frac{k+1}{2}\right\rfloor}{k+1\choose n}{k+1-n\choose n}2^{k+1-2n}\,,
\label{ap14}
\end{align}
\end{widetext}
where the last term is
\begin{align}
\sum_{n=0}^{\left\lfloor\frac{k+1}{2}\right\rfloor}{k+1\choose n}{k+1-n\choose k+1-2n}2^{k+1-2n} = {2k+2\choose k+1}\,,
\label{arxiv}
\end{align}
as stated in \cite{G72}. Hence,
\begin{align}
\nonumber\lefteqn{\sum_{n=0}^{\left\lfloor\frac{k+1}{2}\right\rfloor}{t\choose n}{t-n\choose t-k-1+n}2^{k+1-2n} =}& \\
\nonumber &=  {2t-1\choose k} + \cdots + {2k+3\choose k} + {2k+2\choose k} + {2k+2\choose k+1} \\
 &= {2t\choose k+1}\,,
\label{ap15}
\end{align}
by using (\ref{ap3}) successively. This concludes the proof.


\end{document}